%% file: main.tex
\documentclass[conference]{IEEEtran}

\usepackage{amsmath,amssymb,amsfonts}
\usepackage{algorithmic}
\usepackage{textcomp}
\usepackage{xcolor}
\usepackage{hyperref}
\usepackage{svg}
\usepackage{cite}
\usepackage{amsmath,amssymb,amsfonts}
\usepackage{algorithmic}
\usepackage{algorithm,multirow}
\usepackage{textcomp}
\usepackage{xcolor}
\usepackage{graphics}
\usepackage{complexity}
\usepackage{comment}
\usepackage{multicol}
\usepackage[autostyle, english = american]{csquotes}
\MakeOuterQuote{"}
\usepackage{hyperref}
\usepackage{caption}
\usepackage{subcaption}

\def\BibTeX{{\rm B\kern-.05em{\sc i\kern-.025em b}\kern-.08em
    T\kern-.1667em\lower.7ex\hbox{E}\kern-.125emX}}

\usepackage{graphicx} 
\usepackage{bmpsize}

\begin{document}

\title{QuGAN: A Quantum State Fidelity based Generative Adversarial Network}

\author{
    Samuel A. Stein,\textsuperscript{\rm 1,4}
    Betis Baheri,\textsuperscript{\rm 2}
    Daniel Chen,\textsuperscript{\rm 3}\\
    Ying Mao,\textsuperscript{\rm 4} Qiang Guan,\textsuperscript{\rm 2} Ang Li,\textsuperscript{\rm 1} Bo Fang,\textsuperscript{\rm 1} and Shuai Xu\textsuperscript{\rm 3} \\

    \textsuperscript{\rm 1}Pacific Northwest National Laboratory (PNNL), \{samuel.stein, ang.li, bo.fang\}@pnnl.gov\\
    \textsuperscript{\rm 2}Department of Computer Science, Kent State University, \{bbaheri, qguan\}@kent.edu\\
    \textsuperscript{\rm 3}Computer and Data Sciences Department,Case Western Reserve University, \{txc461, sxx214\}@case.edu \\
    \textsuperscript{\rm 4} Computer and Information Science Department, Fordham University, \{sstein17,ymao41\}@fordham.edu\\

 }

\maketitle

\begin{abstract}
Tremendous progress has been witnessed in artificial intelligence where neural network backed deep learning systems have been used, with applications in almost every domain.
As a representative deep learning framework,  Generative Adversarial Network (GAN) has been widely used for generating artificial images, text-to-image or image augmentation across areas of science, arts and video games.
However, GANs are computationally expensive, sometimes computationally prohibitive. Furthermore, training GANs may suffer from convergence failure and modal collapse. 
Aiming at the acceleration of use cases for practical quantum computers,
we propose QuGAN,  a quantum GAN architecture that provides stable convergence, quantum-states based gradients and significantly reduced parameter sets.
The QuGAN architecture runs both the discriminator and the generator purely on quantum state fidelity and utilizes the swap test on qubits to calculate the values of quantum-based loss functions.
Built on quantum layers, QuGAN achieves similar performance with a 94.98\% reduction on the parameter set when compared to classical GANs. 
With the same number of parameters, additionally, QuGAN outperforms state-of-the-art quantum based GANs in the literature providing a 48.33\% improvement in system performance compared to others attaining less than 0.5\% in terms of similarity between generated distributions and original data sets. QuGAN code is released at \url{https://github.com/yingmao/Quantum-Generative-Adversarial-Network}
\end{abstract}

\begin{IEEEkeywords}
Quantum State Fidelity, Quantum Generative Adversarial Network, IBM-Q
\end{IEEEkeywords}

\input{sections/introduction}
\input{sections/background}
\input{sections/design}

\input{sections/result}

\input{sections/discussion}

\section*{Acknowledgements}
This material is based upon work partially supported by the U.S. Department of Energy, Office of Science, National Quantum Information Science Research Centers, Co-design Center for Quantum Advantage (C2QA) under contract number DE-SC0012704. We thank the IBM Quantum Hub to provide the quantum-ready platform for our experiments. Dr. Mao thanks the Faculty Fellowship from Fordham University.  

\bibliographystyle{IEEEtran}
\bibliography{main}

\end{document}

%% file: sections/introduction.tex
\section{Introduction}
The topic of Generative Adversarial Networks (GANs) has been of significant interest since its discovery \cite{goodfellow2014generative}. GANs have proven to be an exceptionally successful framework for generative modelling, with GANs being able to generate completely unique synthesized samples based off of a real data set. Among the countless applications of GANs \cite{wang2019generative,arjovsky2017wasserstein,dong2018musegan}, one domain where significant achievement has been witnessed is within computer vision. A few common examples within Computer Vision GANs include generating similar images to $18^{\text{th}}$ century artist's paintings \cite{2017cans} to synthesizing images of human faces \cite{NIPS2017_6612}. The application of GANs is not only restricted to the generation of never-before-seen synthesized samples, with applications extending even to domains such as image enhancement techniques, generating higher resolution images from originally low-resolution samples~\cite{8099502}. GANs have been a tremendously interesting but challenging topic within Machine Learning \cite{kumar2020generative}. It pushes the boundary of computational creativity, with the demonstrations becoming more jaw-dropping by the year \cite{Liao2020CVPR}. 

Although GANs have been both tremendously powerful and interesting, their performance is bounded by multiple factors: training a GAN commonly suffers from a stable convergence problem \cite{roth2017stabilizing}, ever-increasing computational complexity in relation to deep neural networks \cite{orponen1994computational} and computer vision tasks, vanishing gradients, and mode collapse \cite{arjovsky2017wasserstein}.  
As contributions from both industry and academia continue to attempt to tackle aforementioned problems, studies on developing Quantum Deep Learning models \cite{garg2020advances,wiebe2014quantum,PhysRevResearch.2.033212,Hueaav2761, stein2022quclassi, stein2021hybrid} have started to draw attention in recent years. This can be attributed most likely due to the allure of quantum supremacy, recently demonstrated by Google \cite{quantumsupremacy}, as well as hopes to find a possible quantum advantage within the machine learning domain. Currently, most Quantum GAN's proposed discuss potential uses such as data loading or attempt to attain an advantage over their classical counterparts \cite{lloyd2018quantum,hu2019quantum,zoufal2019quantum,verdon2019learning,beer2020training,chen2019variational,dallaire2018quantum}. Quantum Deep learning continues to draw attention within Quantum Machine Learning, and has shown signs of significant promise with accuracy's super-seeding similarly designed classical networks \cite{jiang2020can}. This line of Quantum-enabled research is motivated by the aforementioned allure of a potential quantum advantage which could partially alleviate the computational complexity issue, as well as searching for a solution to the mode collapse, growing computational costs of classical neural networks and vanishing gradients problem. An example of an established Quantum GAN is demonstrated in ~\cite{zoufal2019quantum} which makes the use of $O(\text{poly}(n))$ quantum bits to train a GAN to learn random distributions, generating similar patterns.

In this paper, we present the design and implementation of QuGAN, a Quantum GAN architecture.
The key feature of the QuGAN architecture is that the QuGAN model is among the first efforts that run the Discriminator and the Generator on quantum-state based loss functions. With quantum fidelity measurements, these loss functions are enabled by the swap test, a procedure in quantum computation that measures how much two quantum states differ that often comes up in Quantum Machine Learning applications \cite{aimeur2006machine}. As a result, compared to the quantum discriminators and quantum generators proposed in the prior work~\cite{zoufal2019quantum,tfquantum}, our architecture proves to be more stable and performant.
When evaluated with MNIST dataset~\cite{lecun2010mnist}, QuGAN leads to a significant improvement in the Hellinger distance,  which represents the similarity between original dataset and generated probability distributions.
For a comprehensive evaluation, we conduct both simulations on local servers and experiments on real quantum computers.
The key contributions are summarized below.

\begin{itemize}
    \item Based on quantum fidelity measurements, we propose quantum-state based loss functions with quantum gradients for both the Discriminator and the Generator for the GAN models.
    
    \item We evaluate QuGAN on the MNIST~\cite{lecun2010mnist} dataset with PCA~\cite{maaten2008visualizing} reduced dimensions. This extends the application of the prior GAN models on broader domains. 
    
    \item Comparing to Tensorflow based classical GANs, QuGAN achieves nearly identical performance with 94.98\% fewer parameters. Moreover, QuGAN provides a more stable trend to convergence when compared with the state-of-the-art quantum based GANs~\cite{zoufal2019quantum, tfquantum}: we observe a  48.33\%  reduction of the variance versus the 0.5\% reduction for TFQ-GAN and no clear convergent trend for Qi-GAN over the increasing number of the trained epochs.
    
    
\end{itemize}

%% file: sections/background.tex
\section{Background}
A Generative Adversarial Network (GAN) broadly refers to an architecture comprised of two models which compete against one and other, playing an adversarial game. Each model is tasked with attempting to fool the other in a game of cat and mouse \cite{goodfellow2014generative}. The common example of the generator being a fraudster producing fake money (generator), wanting to fool a detective (discriminator) who is trained in discerning between fake and real money. The GAN architecture typically consists of a real data set $(x)$, a Discriminator (D) and a Generator (G). The discriminator's objective is to be able to correctly classify between real and fake data synthesized by G. The generator's objective is to synthesize samples that the discriminator will classify as real. 


With this architecture being rather generic, GANs are agnostic of the type of data they are fed and have applications in an extensive set of domains. However, the performance within each respective application is strongly dictated by the network architecture \cite{wang2019generative}. GANs have shown exceptional results in, but not limited to, computer vision \cite{wang2019generative} and time-based domains such as music composition \cite{dong2018musegan}. In traditional generative networks, G samples from a specific distribution $p_z$ to synthesize a fake sample which is to be fed to D. D is fed with both real data $x$ and data synthesized by G. The discriminator's output is a perception on the data that is fed, and its goal is to maximize $D(x)$  and to minimize $D(G(p_z))$. While the generator's goal is to maximize $D(G(p_z)$. Where probabilistic models aim to label real data as 1 and fake data as 0, the primal min-max optimization formulation is described in Equation \ref{eqn:lossfunc}.

\begin{equation}
\begin{split}
    min_G max_D V(D,G) = \mathbf{E}_{x\text{~}{p}_{data(x)} }[log(D(x)] + \\ \mathbf{E}_{z\text{~}{p}_z(z)}[log(1-D(G(z))]
\end{split}
\label{eqn:lossfunc}
\end{equation}

The one direction in relation to the GAN-related research is to explore the potential of quantum GANs \cite{huang2020experimental,hu2019quantum,lloyd2018quantum}. The conceptual translation between a classical GAN and a quantum GAN is relatively intuitive, as the generator network's task is to learn the underlying probability distribution of a real data set and quantum computers produce probability distributions instead of discrete values. Prior studies propose the use of quantum GANs ranging from for attempting to attain a quantum speed up, or for data loading onto quantum computers 
\cite{zoufal2019quantum,beer2020training,chen2019variational,verdon2019learning}. For example, Zoufal et al.~\cite{zoufal2019quantum} present Qi-GAN to learn random distributions for financial applications. Their work, however, is only extended to one dimensional data learning, and displays significant instability within their models evaluation. Furthermore, they make use of a classical discriminator, and hence making their Qi-GAN only a partial quantum GAN that might not be fully accelerated by quantum systems. Another example of a Quantum GAN~\cite{dallaire2018quantum} makes the use of variational circuits. The authors present a Quantum GAN architecture. However, their learning is rather sporadic and inconsistent, with a non-convergent and  unclear trend. 

Based on the quantum state fidelity based loss functions for Generator and Discriminator,  we propose QuGAN that  provides a more stable and scalable architecture of quantum GANs. 

\subsection{Quantum Fidelity Measurement}

One crucial component regarding the proposed Quantum GAN (QuGAN) architecture is the SWAP test, a quantum state fidelity algorithm. Two states, $|\phi\rangle$ and $|\psi\rangle$, are passed to a system where the measured output is their fidelity. One anicilla qubit in state $\frac{1}{\sqrt{2}}|0\rangle + \frac{1}{\sqrt{2}}|1\rangle$ is passed to the system along side the aforementioned states. The quantum state is described in Equation~\ref{eqn:initial_swap_test_state}.

\begin{equation}
    State = \frac{1}{\sqrt{2}}|0,\phi,\psi\rangle + \frac{1}{\sqrt{2}}|1,\phi,\psi\rangle
    \label{eqn:initial_swap_test_state}
\end{equation}

Following this, the algorithm makes use of a control swap gate, where if the control qubit measures 1 the states are swapped, otherwise nothing happens. The state of $|\psi\rangle$ and $|\phi\rangle$ are control swapped with the anicilla qubit as the control qubit, resulting in the state described in Equation~\ref{eqn:mid_swap_test_state}.

\begin{equation}
\begin{split}
        State = \frac{1}{\sqrt{2}}|0,\phi,\psi\rangle + 
        \frac{1}{\sqrt{2}}|1,\psi,\phi\rangle
    \label{eqn:mid_swap_test_state}
    \end{split}
\end{equation}

The anicilla qubit is passed through a Hadamard gate referenced in Equation \ref{eq:had}, placing the system in the state described in Equation \ref{eqn:post_hadamard}.

\begin{equation} 
H=\left[\begin{array}{cc}
\frac{1}{\sqrt{2}} & \frac{1}{\sqrt{2}} \\
\frac{1}{\sqrt{2}} & -\frac{1}{\sqrt{2}} 
\end{array}\right]
\label{eq:had}
\end{equation}

\begin{equation}
\begin{split}
        State = \frac{1}{2}(|0,\phi,\psi\rangle + |1,\phi,\psi\rangle +
        |0,\psi,\phi\rangle - |1,\psi,\phi\rangle 
 \end{split}
\label{eqn:post_hadamard}
\end{equation}

To attain the probability of measuring the anicilla qubit as 0, we square sum the states where $|0\rangle$ is measured.

\begin{equation}
\begin{split}
    State = \frac{1}{2}|0\rangle(|\phi\rangle|\psi\rangle + |\psi\rangle|\phi\rangle) +  \frac{1}{2}(|\phi\rangle|\psi\rangle - |\psi\rangle|\phi\rangle)
 \end{split}
    \label{eqn:state_based}
\end{equation}

To attain Equation~\ref{eqn:final_probability_swap}, we isolate the coefficients to the $|0\rangle$ state and square them to attain a P(Measurement=0), as seen in Equation \ref{eqn:state_based}. As described in Equation \ref{eqn:final_probability_swap},  the fidelity of states $|\phi\rangle$ and $|\psi\rangle$ is bound to [0.5,1], and measures the fidelity of two states. In the case of the states being orthogonal, the dot product between the matrices will equal 0 and the fidelity measures 0.5, and if the matrices are identical the dot product measures 1 and the fidelity measures 1.

\begin{equation}
    P(Measurement=0) = \frac{1}{2} + \frac{1}{2}|\langle\psi,\phi\rangle|^2
    \label{eqn:final_probability_swap}
\end{equation}

%% file: sections/design.tex
\section{QuGAN System Design}
\label{design}

\begin{figure*}
\centering
         \includegraphics[width=0.75\linewidth]{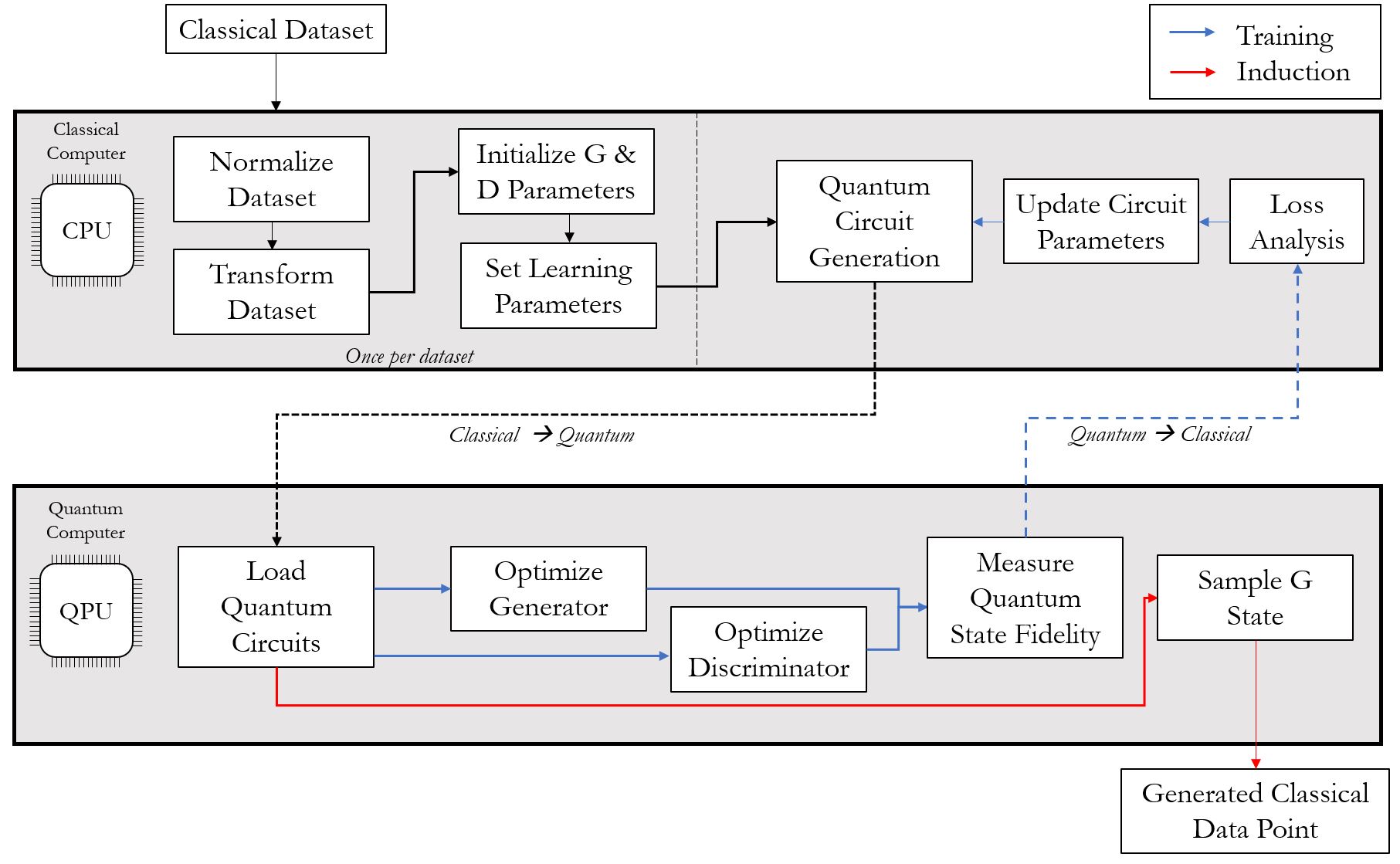}
\caption{QuGAN System Architecture}
      \label{fig:sys}
\end{figure*}

In this section, we introduce the QuGAN system architecture design in details. As shown in Figure \ref{fig:sys}, the system operates through iterative handovers between a classical and quantum computer.  
It operates broadly by the iterated upon operation of a classical computer passing a parameterized quantum circuit to a quantum computer, which then passes back a measured quantum state fidelity. QuGAN architecture begins with being fed classical data, which is normalized and transformed into quantum data. This data-prepossessing step is done once per data set. A Generator/Discriminator circuit or Real Data/Discriminator circuit is passed to the quantum circuit, where the circuit passed is chosen based on the stage of the training algorithm. The quantum circuits of induced state fidelity's are transferred back to the classical computer. In the case of the system being optimized, the fidelity is used in the calculation of each gates gradient with respect to the objective function, which is used to update the Generator and Discriminator parameters. If the system is being used to generate samples, instead of passing back a state fidelity the quantum computer is sampled from and a data point generated.
QuGAN system is iterated upon a specific number of times, or until a certain goal has been reached.


In existing Quantum generative research, models have looked at taking a traditional Discriminator but changing the Generator to a quantum Generator \cite{chen2019variational}, or even implementing a Quantum Wasserstein GAN~\cite{wgan}. The problem with the Quantum/Classical approach however is that a classical generator fails tackle quantum data efficiently, and requires the quantum data to be translated back to classical data. Furthermore, in the case of the Quantum Wasserstein GAN, this makes use of Earth Movers distance over a density matrix. This approach, although valid, is not as applicable to QuGAN architecture as our quantum state fidelity is measured through a single ancilla qubits expectation value that captures the similarity between two quantum states. One of Wasserstein Distance's advantage over classical log likelihood loss functions it that its range is not bound between $0$ and $1$. However, through measuring the expectation of a single qubit, network outputs remain bound even when using Wasserstein distance. Therefore, we deem it appropriate to make use of the log-likelihood approach for optimization. We make use of a Quantum Generator as well as a Quantum Discriminator (details in Equation~\ref{eqn:g_loss_quantum} and \ref{eqn:d_loss_quantum}), and accomplish communication between the models through the SWAP test. The quantum circuits are simulated on a classical computer, with a proof of concept illustration by running the circuits on actual quantum computers.

\subsubsection{Quantum Deep Learning Layers}

\begin{figure}
\centering
\begin{subfigure}{.30\textwidth}
  \centering
  \includegraphics[width=\textwidth]{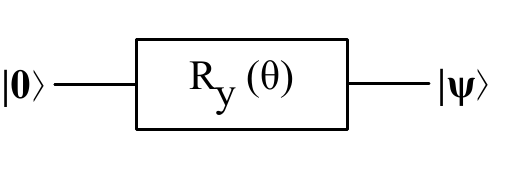}
  \caption{Single-Qubit Unitary}
  \label{fig:sub-first}
\end{subfigure}
\begin{subfigure}{.30\textwidth}
  \centering
  \includegraphics[width=\textwidth]{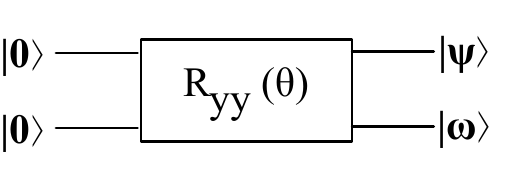}
  \caption{Dual-Qubit Unitary}
  \label{fig:sub-second}
\end{subfigure}
\begin{subfigure}{.30\textwidth}
  \centering
  \includegraphics[width=\textwidth]{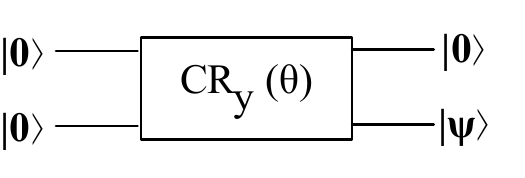}
  \caption{Entanglement Unitary}
  \label{fig:sub-third}
\end{subfigure}
\caption{Three Types of Parameterized Quantum Layers}
\label{fig:squ}
\end{figure}

The modelling of deep learning on a quantum system is commonly represented by collections of quantum gates which are grouped together such that they accomplish specific quantum data manipulation \cite{chen2019variational}. Each parameterized gate accomplishes a specific type of quantum state manipulation \cite{divincenzo1998quantum}. We adopt this variational quantum circuit approach and expand it to create 3 key types of gate combinations, a single qubit unitary, dual qubit unitary, and an entanglement unitary visualized in Figure \ref{fig:squ}. A single qubit unitary, as seen in Figure \ref{fig:squ}(a), performs an $R_Y$ gate on a single qubit, parameterized by $\theta$. This gate accomplishes a manipulation of a single qubits superposition. The dual qubit unitary, drawn in Figure \ref{fig:squ}(b),  performs a rotation on 2 qubits around the Y-axis, parameterized by one $\theta$. This gate accomplishes a manipulation on two qubits superposition. Accomplishing  entanglement, QuGAN makes use of an entanglement unitary using a $CR_Y(\theta)$ gate parameterized by one $\theta$, as plotted in Figure \ref{fig:squ}(c). The $CR_Y(\theta)$ gate rotates a qubit by $\theta$ provided the control qubit measures 1, entangling the qubits.
These grouped gate operations all accomplish specific quantum data manipulation. The gates unitary matrices are outlined in Equations \ref{eq:ry}, \ref{eq:ryy} and \ref{eq:cry}.

\begin{figure*}
\centering
         \includegraphics[width=0.9\linewidth]{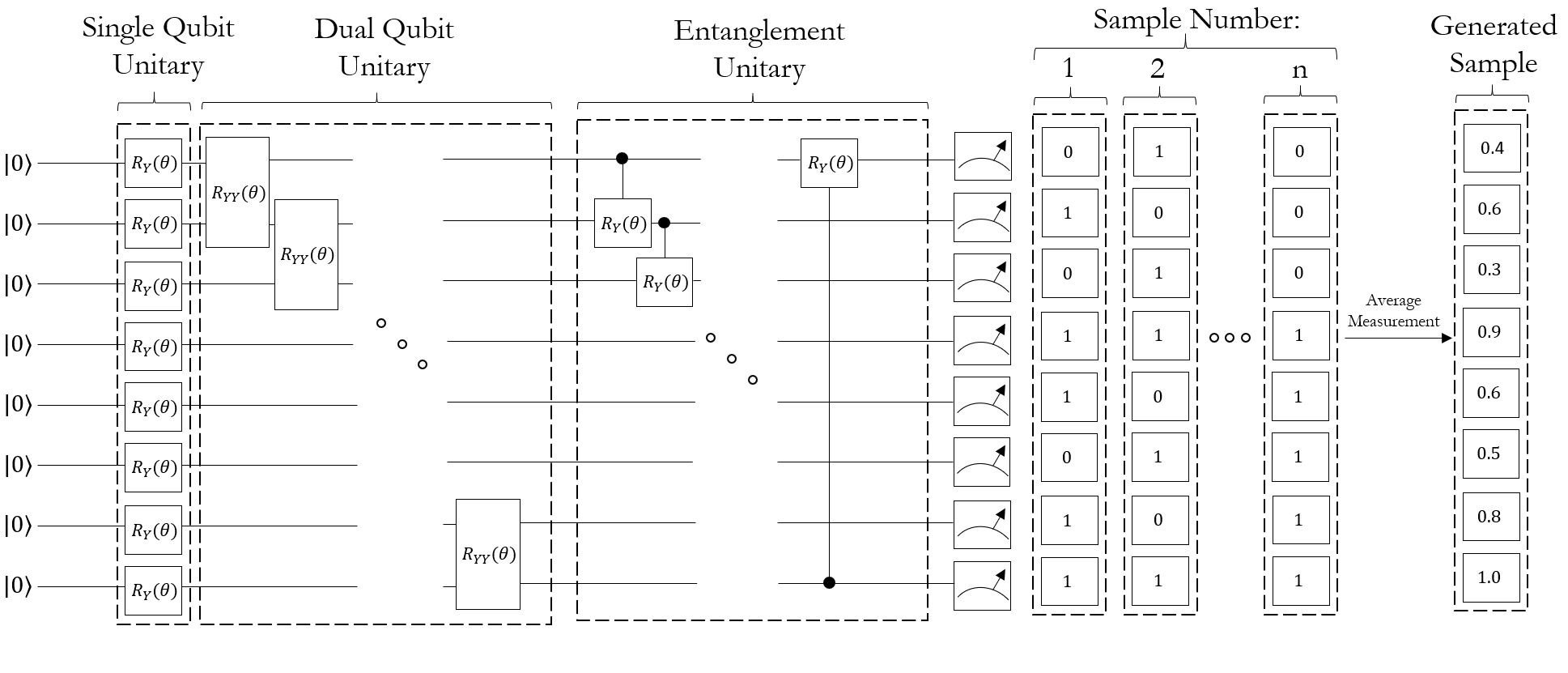}
\caption{Layered Quantum Network Design}
      \label{fig:layered_design}
\end{figure*}

The operations visualized in Figure \ref{fig:squ} are grouped together to form a layer of operations on the quantum state parameterized by $\theta_d$. Within the Figure~\ref{fig:layered_design}, we illustrate each of these layer types in order as well as illustrate how a circuit can be used to generate classical numerical outputs. Which qubits are measured and how this output relates to the optimization function is up to the practitioner. In the case of a Generator, the output sample would be a synthesized data point.

\begin{equation} 
R_{Y}(\theta)=\left[\begin{array}{cc}
\cos \left(\frac{\theta}{2}\right) & -\sin \left(\frac{\theta}{2}\right) \\
\sin \left(\frac{\theta}{2}\right) & \cos \left(\frac{\theta}{2}\right)
\end{array}\right]
\label{eq:ry}
\end{equation}

\begin{equation}
R_{YY}(\theta)=\left[\begin{array}{cccc}
\cos{\frac{\theta}{2}} & 0 & 0 & i\sin{\frac{\theta}{2}} \\
0 & \cos{\frac{\theta}{2}}  & -i\sin{\frac{\theta}{2}} & 0 \\
0 & -i\sin{\frac{\theta}{2}} & \cos{\frac{\theta}{2}}  & 0 \\
i\sin{\frac{\theta}{2}} & 0 & 0 & \cos{\frac{\theta}{2}}
\end{array}\right]
\label{eq:ryy}
\end{equation}

\begin{equation}
CR_{Y}(\theta)=\left[\begin{array}{cccc}
1 & 0 & 0 & 0 \\
0 & 1 & 0 & 0 \\
0 & 0 & \cos{\frac{\theta}{2}} & -\sin{\frac{\theta}{2}} \\
0 & 0 & \sin{\frac{\theta}{2}} & \cos{\frac{\theta}{2}} 
\end{array}\right]
\label{eq:cry}
\end{equation}

\begin{figure*}[!htb]
\centering
         \includegraphics[width=0.85\linewidth]{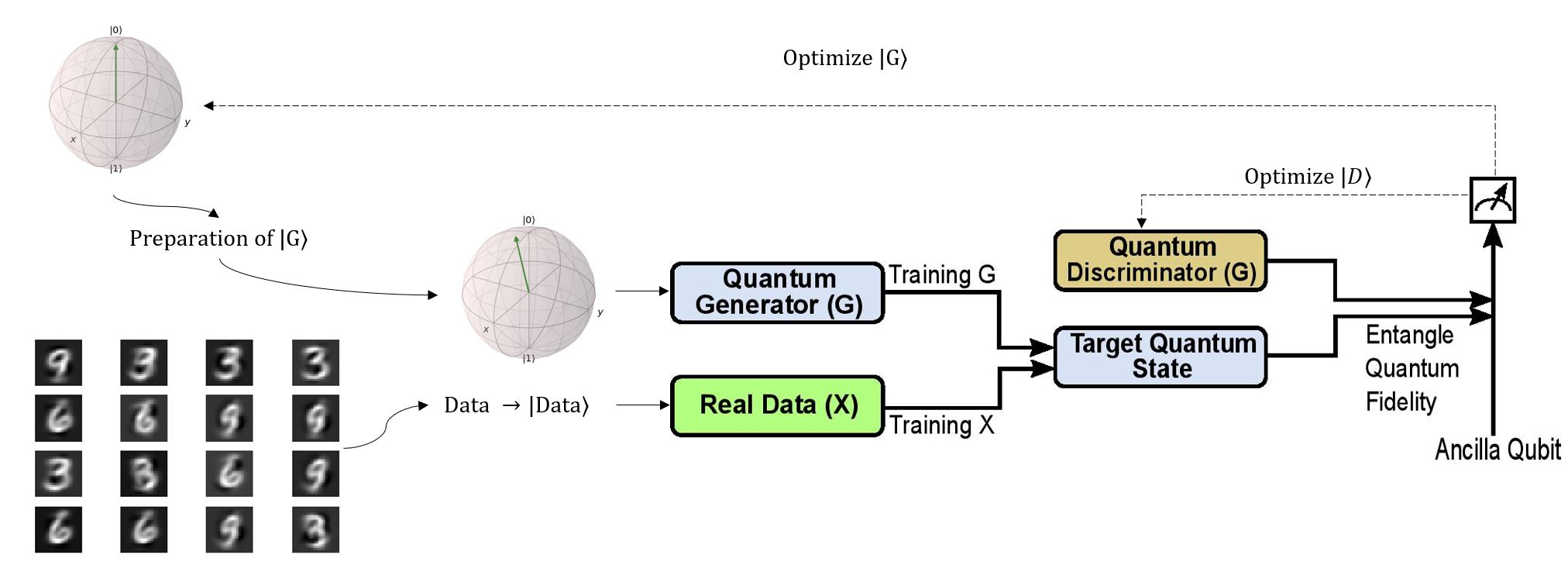}
\caption{Inter-model Communication Architecture}
      \label{fig:comm_arch}
\end{figure*}

For QuGAN, the architecture is comprised of a Discriminator (a Quantum based Neural Network), the Generator (another Quantum based Neural Network), real data and an anicilla qubit to measure Quantum fidelity. We design both Discriminator and Generator equally to allow for equal ability in learning of the ideal quantum state. The general architecture of Our QuGAN on the actual quantum computer, described in the "Quantum Computer" block of Figure \ref{fig:sys}, is outlined in Figure \ref{fig:comm_arch}. 

In Figure \ref{fig:comm_arch}, we illustrate the Quantum Computer's role through the two possible styles of circuits being prepared. In the case of $|G\rangle$, the circuit prepares the two respective states using the layered approach. If there is data being prepares, only the Discriminator makes use of a layered approach, whilst the data is encoded using $R_Y(\theta)$ gates, which is fed onto a circuit along side the Discriminator.

The generated quantum circuit operates by whether the system is measuring the Discriminator (D/$|\delta\rangle$) loss relative to real data or fake data, which is illustrated by the "Generator" (G/$|\gamma\rangle$) and "Real Data" (X/$|\xi\rangle$) blocks, based on whether we the system is measuring D/G or D/X loss. The discriminators loss in terms of quantum states is described in Equation \ref{eqn:d_loss_quantum} and generators in Equation \ref{eqn:g_loss_quantum}. The terms within the log relate to the quantum fidelity measurement being normalized to a [0,1] scale, hence why there is no $\frac{1}{2}$. To generate a sample, the quantum state representing the Generator is induced and sampled from $n$ times, and is visualized in Figure \ref{fig:sys}. The average measurement per each qubit represents a specific dimensions output. 

\begin{equation}
\begin{split}
    D_{loss}=\mathbf{E}[log(D(|\langle \xi,\delta \rangle|^2] + \\
    \mathbf{E}[log(1-D(|\langle \gamma,\delta \rangle |^2]
    \label{eqn:d_loss_quantum}
\end{split}
\end{equation}
\begin{equation}
    G_{loss} = \mathbf{E}[log(D(|\langle \gamma,\delta \rangle |^2]
    \label{eqn:g_loss_quantum}
\end{equation}

The gradient of the loss function is used to update the generator and discriminators states respectively to improve their performance in the adversarial game.


\subsubsection{Quantum based QuGAN Optimization}

Optimization of a Quantum GAN can be characterized by the loss function described in Equation \ref{eqn:d_loss_quantum} and \ref{eqn:g_loss_quantum} . The design of our QuGAN is comprised of $4$ key quantum circuit components, namely the Generator Circuit, Discriminator Circuit, Data Loading Circuit and the Anicilla qubit. The Generator circuit represents the quantum deep learning (QDL) model tasked with generating samples; the discriminator represents the QDL model that is tasked with discerning between fake and real samples; the data loading circuit is responsible for loading quantum data onto a quantum state; the anicilla qubit is a single qubit with which we measure quantum state fidelity and through which we accomplish our inter-circuit communication. This single qubit being used to measure fidelity has its benefits within the quantum setting, as near term devices can be rather noisy \cite{preskill2018quantum,xue2021effects} and hence using lower qubit counts requiring measurement can reduce the number of samples needed to attain a confident reading from the quantum computer. Therefore, we find this to be one of the advantages to our model such that we are only measuring one qubit to encapsulate our entire systems performance. A range of $[0.5,1]$ would not necessarily work as well within a log optimization problem, therefore we normalize our SWAP test value to be between $[0,1]$, which is seen in the final equation forms in \ref{eqn:d_loss_quantum} and \ref{eqn:g_loss_quantum}. This possible range of measurements is further motivation to use the original GAN formulation of a log loss function instead of other approaches such as Wasserstein distance. In comparison and further justifying our use of a probabilistic loss, Wasserstein distance makes use of a linear output neuron in classical GANs, attaining values in the range of $[-\infty,\infty]$, one of the large advantages of Wasserstein GANs, where quantum state observables are bound by $[0,1]$.

These quantum state fidelity's are used to calculate both the Discriminator and Generator loss. We can update the parameters of the networks similarly to a classical GAN, as visualized in Figure \ref{fig:sys}, through the use of gradient descent and quantum gate differentiation. In differentiating a quantum gate, we attain the gradient of a parameter in the gates layers weight vector. The system proceeds by measuring the loss of the Discriminator with respect to both the Generator and the Real Data, then updates the quantum deep learning model to improve the performance at a rate of $\alpha$ (learning rate). Following this the generator analyzes how well it performs against the Discriminator and updates its quantum deep learning model to improve performance respectively. To perform gradient descent of each gate, we employ a parameterized differential equation from \cite{crooks2019gradients} outlined in Equation \ref{eqn:dff}.  

\begin{equation}
    \frac{\delta{f}}{\delta{}\theta{}} = \frac{1}{2}({f(\theta + \frac{\pi}{2}) - f(\theta - \frac{\pi}{2}))}
    \label{eqn:dff}
\end{equation}

Where in Equation \ref{eqn:dff}, $\delta\theta$ indicates a specific $\theta$ gradient, and $f$ is the cost function for the network, which in our case is described in Equation \ref{eqn:d_loss_quantum} and \ref{eqn:g_loss_quantum}. This approach allows for analytical differentiation of the Quantum Neural Network with the downfall being that this approach is computationally expensive having to induce the network twice per gate to attain the gradient.

\subsubsection{Data Qubitization and QuGAN Algorithm}
Data encoding on quantum computers continues to be an area of research, however is not a primary focus of this paper. Hence we propose our method, but acknowledge alternate methods exist and could be tested on our architecture. Taking a classical numerical data set of values $x_1,x_2,...,x_n$ with dimensions $d$, stored through the use of classical bits, using data types such as floats or integers. We note the limitation of  qubit counts renders this method of data encoding infeasible, and additionally this method does not make use of the potential of the exponential Hilbert space representing quantum states. In designing our data encoding method,  we operate on the premise of reversibility, which allows for the transformation of classical data into quantum data, and then returned back into its classical version. The $\mathbf{E}$ value of a qubit can be used to encode classical data, however the expected value range is between 0 and 1. Therefore, we normalize each dimension $d_i$ range to be between 0 and 1. Following this, we encode a data point of value $x$ to a rotation $\theta = 2\sin^{-1}({\sqrt{x}})$. This results in the $\mathbf{E}$ value of a qubit equating to the classical data point.

The Algorithm~\ref{alg:1} illustrates the process we go through to implement our design. In Line 1 we initialize all of our initial parameters for the Generator and the Discriminator. The network weights line indicates the initialization of the parameter vector for both models. Each entry is initialized as a randomly sampled value from a uniform distribution of bounds $[0,\pi]$. We set our epoch count, load our data set, and indicate the ratio of which we will train the discriminator against training the generator. Proceeding this, from Line 2, we begin training our model. We train the discriminator on the real data set once in Lines 4 through 6, followed by training the discriminator on fake samples from the generator in Lines 7 through 10, and finally train the generator on the discriminator in a total of $R$ times (Line 11 - 14).

\begin{algorithm}[!htb]
\caption{Quantum States based Learning}
\label{alg:1}
\begin{algorithmic}[1]
\STATE Parameter Initialization: 
		\par \quad $\text{Learning Rate}: \alpha  = 0.01$
		\par \quad $\text{Network Weights:}$ 
		$\theta_d = \text{Random(0, 1)}\times\pi$
		\par \quad $\text{epochs:} \epsilon= 25$
		\par \quad $\text {Dataset:} X $
		\par \quad $\text {D-to-G Train Count:} I $
		\par \quad $\text {G-to-D Train Count:} R$
		\par \quad $\text {Qubit Channels}: Q = 1 + (n_{X_{dim}}\times2)$
\FOR {$\zeta \in \epsilon$}
    \STATE Train The Discriminator On Real Data
    \FOR {$x_k \in X$}
        \STATE  $Grad = \frac{dCost_{x_k}}{d\theta_D}$
        \STATE Update $\theta_D \xleftarrow[]{}\theta_D - \alpha Grad$
    \ENDFOR
    \STATE Train the discriminator on the generator
    \FOR {$i\in I$}
        \STATE  $Grad = \frac{dCost_{D}}{d\theta_D}$
        \STATE Update $\theta_D \xleftarrow[]{}\theta_D - \alpha Grad$
    \ENDFOR
    \STATE Train the Generator on Discriminator
    \FOR {$j\in R$}
        \STATE  $Grad = \frac{dCost_{G}}{d\theta_G}$
        \STATE Update $\theta_G \xleftarrow[]{}\theta_G - \alpha Grad$
    \ENDFOR
\ENDFOR
\end{algorithmic}
\end{algorithm}


%% file: sections/result.tex
\section{QuGAN Evaluation}
In this section, we discuss the implementation of QuGAN and experiments that we conducted to evaluate it.


QuGAN is implemented based on Python 3.8 with  IBM Qiskit, an open-source framework for quantum computing. 
The generated quantum circuits are trained on GPU-enabled servers on the Google Cloud Platform. 
In addition, we conduct experiments on IBM-Q Experience to validate QuGAN with real quantum computers.

To produce classical results from the QuGAN, we sample the Generator circuit $n$ times and take the mean of those measurements per qubit. 
In the evaluation, we set $n=30$ to attain a data point. For example, we sample on a Qubit $|0\rangle$ 14 times, and $|1\rangle$ 16 times, therefore giving us the measured value of $\frac{(14(0) + 16(1))}{30} = 0.533$ on said qubit.

For a comprehensive evaluation, we compare QuGAN with the following state-of-the-art solutions in the literature. 

\begin{itemize}
    \item {\bf C-GAN}: a classical GAN, which is implemented and trained with different number of parameters as well as epochs under TensorFlow framework with methods similar to that outlined in \cite{arjovsky2017towards}.
    \item {\bf Qi-GAN}: a quantum GAN~\cite{zoufal2019quantum} that is designed to learn random distributions. It is implemented with Qiskit.
    \item {\bf TFQ-GAN}: a quantum GAN ~\cite{tfquantum} that utilizes the TensorFlow quantum architecture proposed in \cite{dallaire2018quantum}.
\end{itemize}

\begin{figure}[!htb]
    \centering
    \includegraphics[width=1\linewidth]{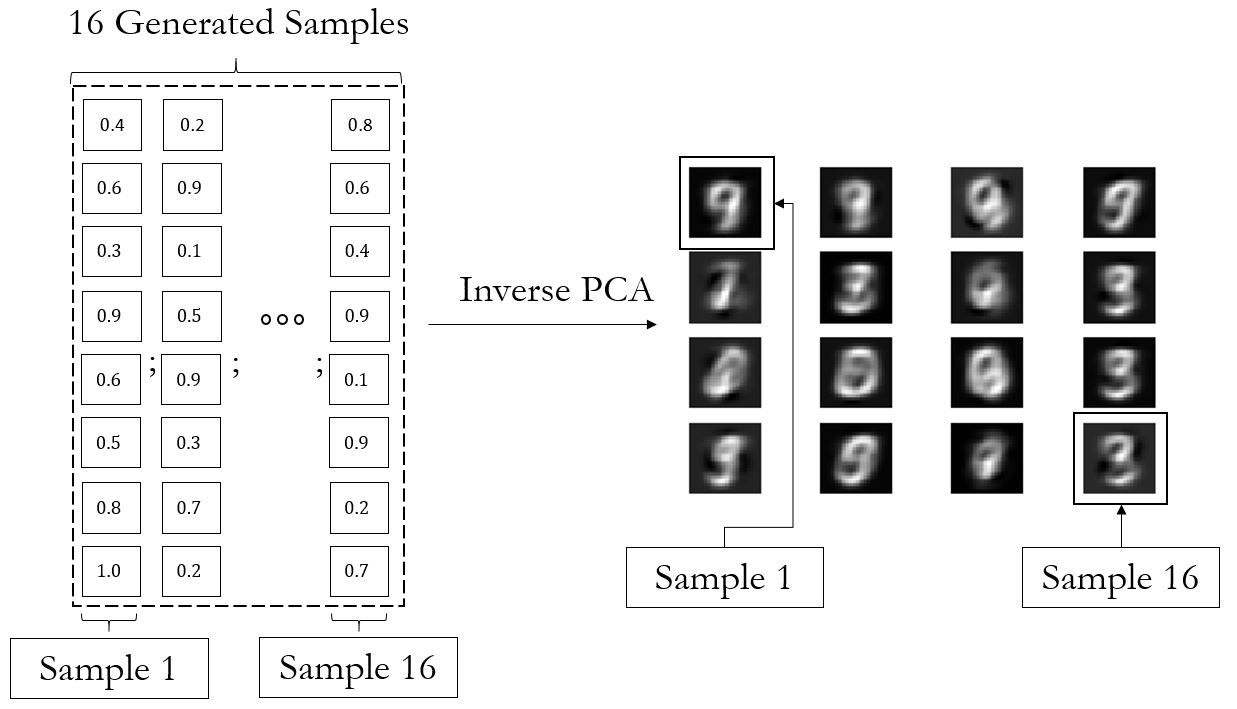}
    \caption{Image Generation Process: To represent our output, a sampling average of each qubit over the Generator is passed through an Inverse-PCA algorithm. This allows us to intuitively visualize and critique our results.}
    \label{fig:sample_generation}
\end{figure}

\begin{figure*}[!htb]
    \centering
    \includegraphics[width=0.8\linewidth]{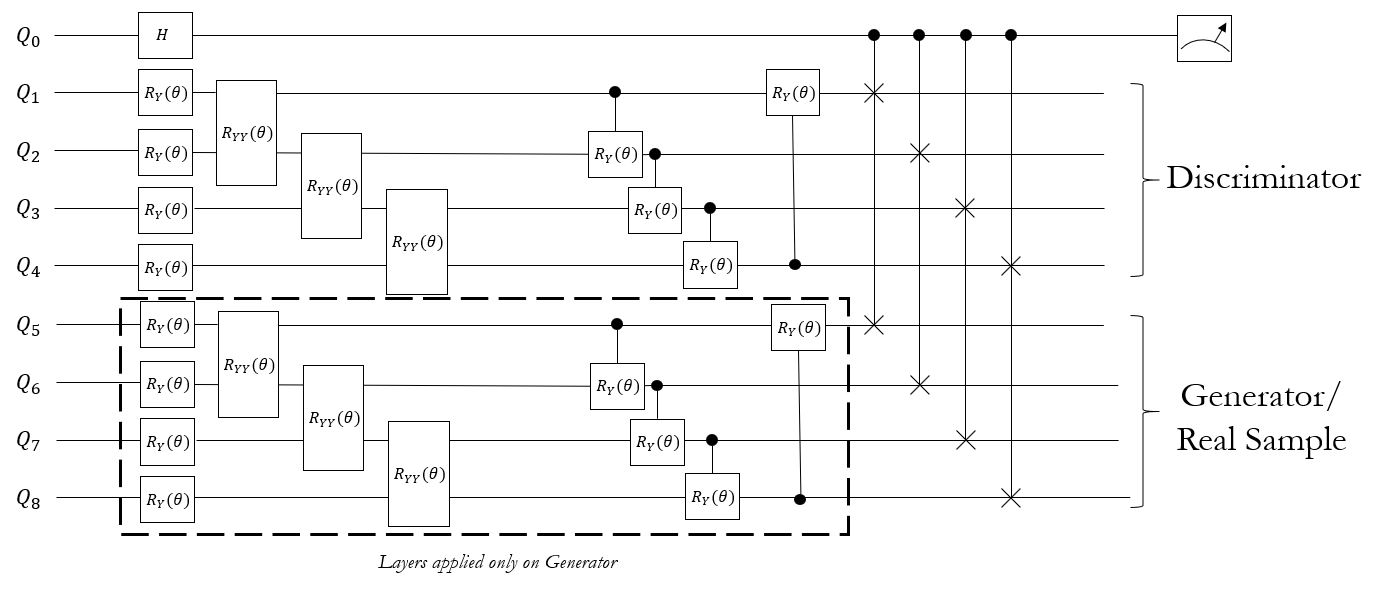}
    \caption{QuGAN Circuit with PCA MNIST data set: Visualized on the circuit, the Discriminator and Generator have 10 parameters ($\theta$ on the figure) each. Qubits 1 through 4 are the Discriminator, and qubits 5 through 8 are reserved for the Generator parameters or data loading. Qubit 0 is the ancilla qubit for the SWAP test. Qubit 0 is measured onto a classical bit which reads either 1 or 0.}
    \label{fig:base_circuit}
\end{figure*}

To evaluate the models, we use {\em Hellinger Distance}~\cite{nikulin2001hellinger} as our key metric to compare the generated distribution and original data sets. Hellinger distance is a popular tool to quantify the similarity between two probability distributions and encompasses the difference in two distributions. 

Although modern classical GAN's are evaluated on qualitative measures, such as Inception Score (IS) \cite{barratt2018note}, this evaluation is not fit for our dataset due to the minimum requirements on dimensions (e.g. limited number of qubits). Furthermore, the use of Hellinger Distance is a well-suited measurement for a quantum space generator, as the output of the Generator is naturally a distribution, which will have a certain distance to the true data. This difference in distributions encapsulated in one number is ideal for evaluating our GAN's performance. If our output space probability distribution stops moving (i.e. gate gradients have vanished ), we will observe no change in the Hellinger Distance. Hence, Hellinger Distance we believe is a good metric to observe for quantum state stability and network convergence.


\subsection{QuGAN: MNIST Learning}
To demonstrate the learning ability of our architecture, we evaluate our Quantum GAN over the MNIST data set. The MNIST data set contains 60,000 images of hand-drawn digits, labeled with their respective class (i.e. an image of a hand-drawn 4 is labelled as a 4). 

The resolution of the images is $28\times28$, which is a high dimensionality for current quantum simulators as well as near-term quantum computers. Therefore, we make use of Principal Component Analysis (PCA)  algorithm to downscale \cite{wold1987principal} the dimensionality from 784 to 4 with an explained variance of $28\%$. To generate images we sample the QuGAN circuit 30 times and take the mean measurement of each qubit, to which we transform the data points generated from the Quantum GAN back through the PCA transformer into 784 dimensions. From here, we draw the images and attain our QuGAN generated images. The use of PCA as a data compression tool is motivated by previous works highlighting the possibility of a quantum PCA algorithm \cite{lloyd2014quantum}. To track the progress of our algorithm, we measure the Hellinger Distance between the Generator and the true data set.  To generate a sample image, we extend the sample generation described Section~\ref{design} into Figure \ref{fig:sample_generation}, whereby 16 samples are generated by our quantum circuit to which we apply Inverse PCA, which visualizes our Generator output. The architecture of our Generator/Discriminator is visualized in Figure \ref{fig:base_circuit}, representing 3 of the 4 key components - namely the Discriminator, Ancilla Qubit, and the Generator. For a Real Sample, instead of $Q_5$ through $Q_8$ having trainable rotations, a data point is loaded instead.



\subsubsection{Comparing to C-GAN} In comparing our architecture to a classical GAN, we run a Tenforflow-based classical GAN with $k$ parameters over $100$ epochs, and measure the average Hellinger Distance using 100,000 samples generated from the generator, trained on classes 3/6/9. As seen, the classical GAN with a parameter count of 20 in the Generator and 20 in the Discriminator performs significantly worse than our QuGAN, with our QuGAN attaining a 56.14\% improvement with respect to Hellinger distance. We increase the parameter count up until 199, where the Hellinger distance begins to approach similar values to our QuGAN, with distances less than $0.3\%$ difference. These results are outlined in Table~\ref{tab:classiccompare}. We illustrate that to attain similar performance in a classical GAN, we need to use 199 parameters versus the 10 in our QuGAN - a 94.98\% reduction in parameters for a similar performance. 

\begin{figure*}[!htb]
\includegraphics[width=1\linewidth]{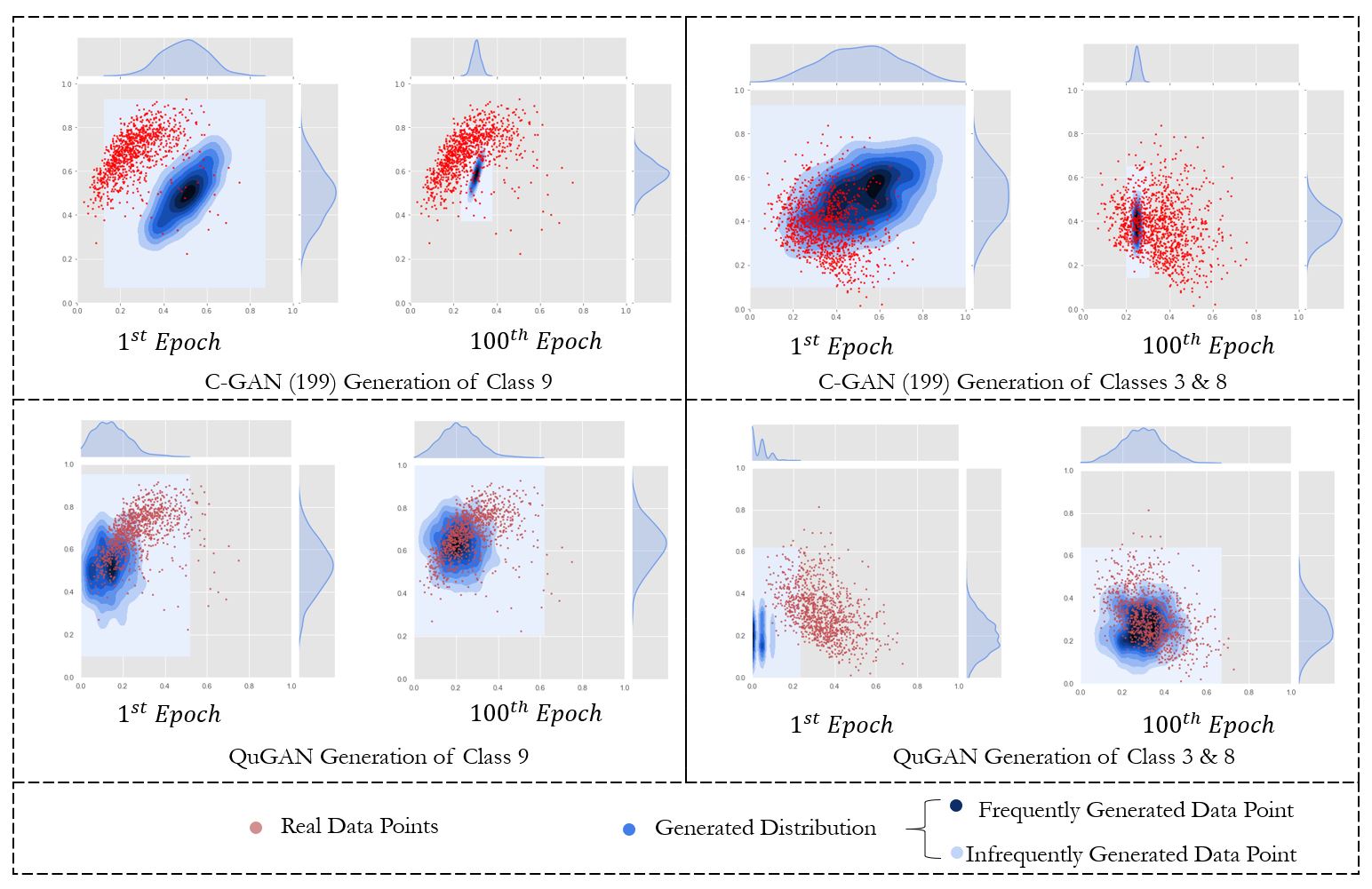}
\caption{Generator Learned Distribution for MNIST: In comparing to a C-GAN (199), our generator is able to attain a significantly more diverse output in comparing to the 2D representation, with both classical experiments consistently leading to mode collapse \cite{arjovsky2017wasserstein,durall2020combating}. We show how our architecture represents a more general and diverse output over its classical counterpart. }
\label{fig:mnist-2d-track}
\end{figure*}

\begin{figure*}[ht]
\includegraphics[width=1\textwidth]{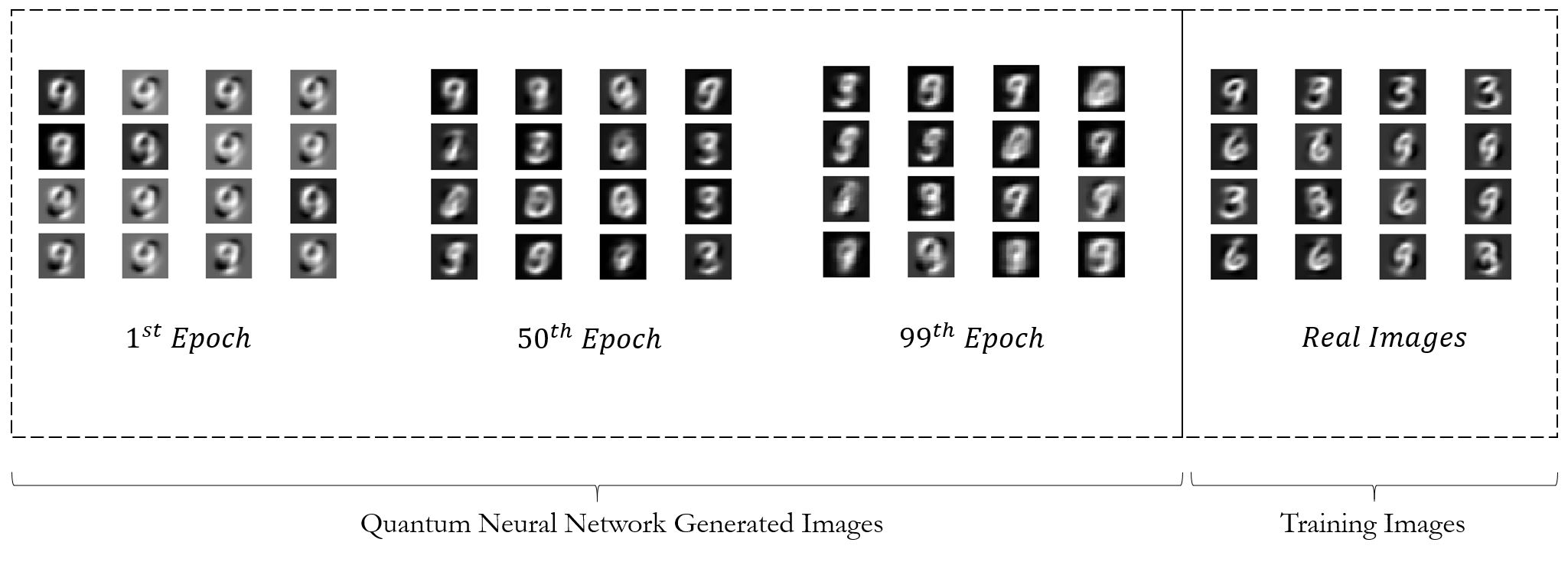}
\caption{QuGAN Generated Images (3/6/9): We illustrate our image generation process over 99 epochs. Critiquing the images generated on the $100^{th}$ epoch, targeting our respective labels we trained the generator on we see strong signs of a 9 being generated in row 2 column 4, evidence of a 6 in row 1 column 4, and a 3 in row 3 column 2}
\label{fig:mnist_images}
\end{figure*}

For a comprehensive evaluation, we further compare a Classical GAN to generating 2D MNIST distributions, as visualized in Figure~\ref{fig:mnist-2d-track}. In this experiment, we illustrate our architecture on the values combinations $\{3,8\},\{9\}$ down scaled to 2-dimensions for the purpose of learning visualization, where we illustrate the output distribution of our Generator (Blue distribution) against the actual data points (Red values). This is visualized in Figure \ref{fig:mnist-3-8-9-track}. Initially, a Hellinger Distance of $0.7561$ is obtained for $\{9\}$ and $0.8157$ for $\{3,8\}$. After 100 epochs, the Generator has converged to a Hellinger distance of $0.08171$ for ${3,8}$ and $0.0875$ for ${9}$. We note that the ideal outputs too are relatively noisy, as with many encoding to decoding (i.e. PCA) processes suffer from image blurriness post decoding. Therefore, our generators success should be compared to the "Real Images" in Figure \ref{fig:mnist_images}.

\begin{table}[t]
\centering
 \begin{tabular}{c c c c }
 \hline
 Architecture & Parameters & Hellinger Distance 
 \\ \hline
 {\bf QuGAN} & 10 & 0.1951 
 \\ 
 \hline \hline
C-GAN & 20 & 0.4448 
\\
 \hline
 C-GAN & 45 & 0.3234 
 \\
 \hline 
 C-GAN & 96 & 0.2515 
 \\ 
 \hline 
 C-GAN & 199 & 0.1945 
 \\
 \hline
 C-GAN & 1299 & 0.1389 
 \\
  \hline
\end{tabular}
\caption{Different GAN Architectures with 50 Epochs}
\label{tab:classiccompare}
\end{table}

\begin{figure}[!htb]
    \centering
    \includegraphics[width=0.9\linewidth]{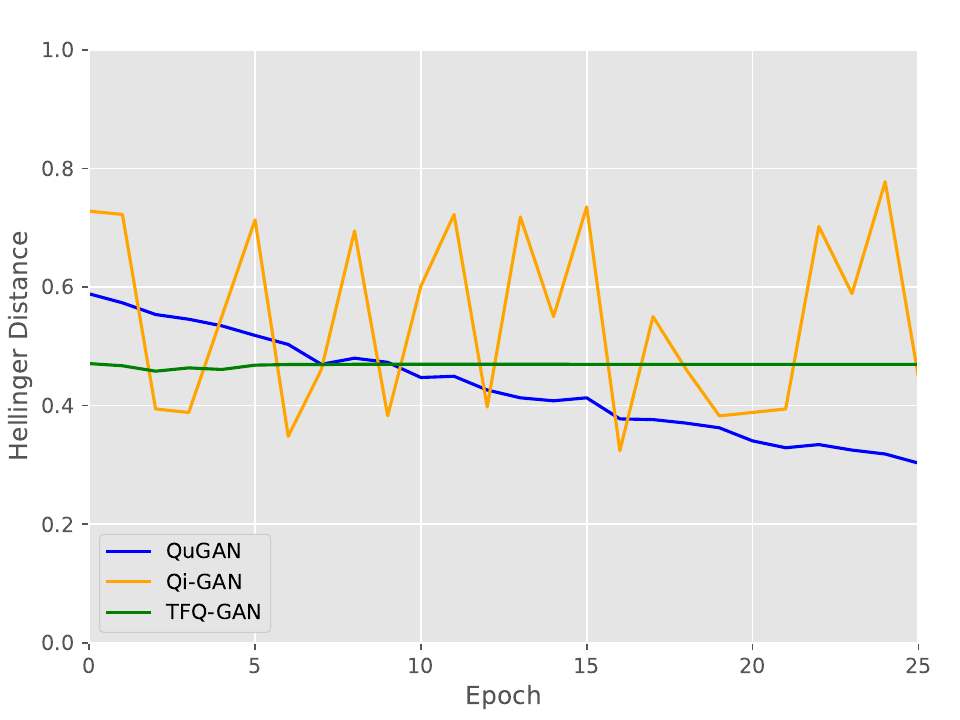}
    \caption{Comparison with TFQ-GAN and Qi-GAN}
    \label{fig:gnd}
\end{figure}

\subsubsection{Comparing to Qi-GAN and TFQ-GAN}
Next, we compare our QuGAN architecture with state-of-the-art quantum-based GANs, Qi-GAN and TFQ-GAN, on the PCA-MNIST dataset $\{3,9\}$ trained over 25 epochs. As visualized in Figure \ref{fig:gnd}
we illustrate QuGAN's stable learning ability by comparing to alternative quantum learning architectures. 
In this figure, we see no signs of consistent learning over Qi-GAN, nor with the TFQ-GAN in terms of Hellinger Distance. Comparing to our architecture, we see QuGAN has a stable consistent convergence. As for Qi-GAN, however, 
the training is extremely sporadic. This sporadic nature is found in our evaluation of the distance. With our architecture, over 25 epochs, a $48.33\%$ reduction in Hellinger Distance was attained, however for TFQ-GAN a less than $0.5\%$ change occurred, and Qi-GAN had no consistent trend and hence any changes attributed to noise.

\subsubsection{Learning Stability} 
\begin{figure}
    \centering
    \includegraphics[width=0.5\textwidth]{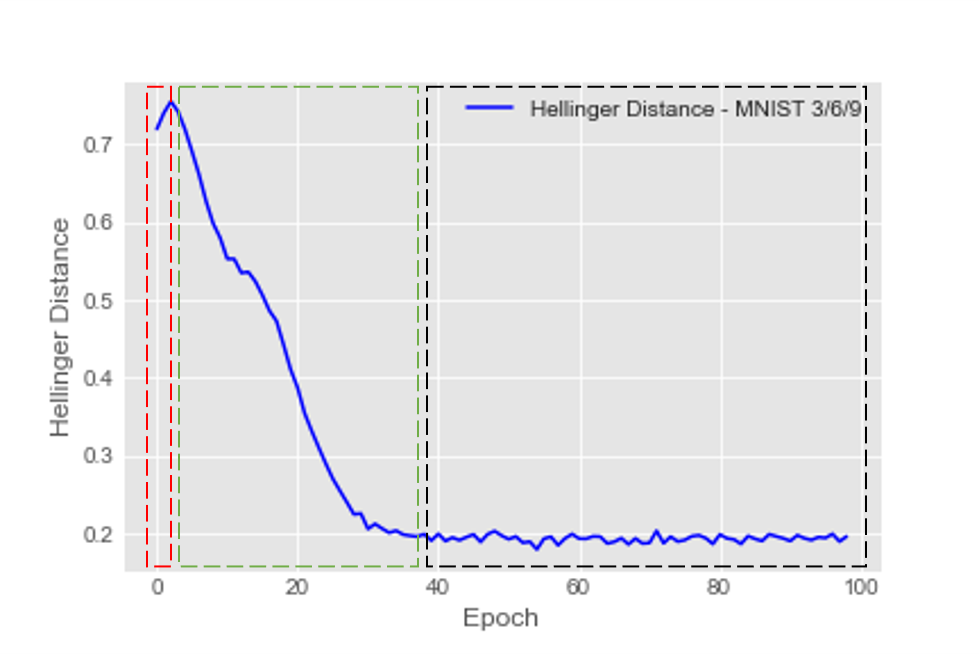}
    \caption{Hellinger Distance of MNIST 3/6/9}
    \label{fig:mnist-3-8-9-track}
\end{figure}

In this experiment, we train the circuit on the images of class 3, 6 and 9 from the MNIST-PCA dataset. The evolution of generated images is visualized in Figure \ref{fig:mnist_images}. As seen in Figure \ref{fig:mnist_images}, initial generated images are unrecognizable with the same noise being generated each time. This consistent noise can be attributed to PCA's focus on preserving dimensions of highest variance. Hence, corners remain black and without interference however the center where the strokes commonly comprise numbers are highly activated. We notice some learning by the $25^{th}$ epoch, with improvements by the $50^{th}$ epoch. This is captured within the Hellinger Distance tracked in Figure \ref{fig:mnist-3-8-9-track} that illustrates our learning process through the three stages of a Generator learning. Initially, the discriminator provides no meaningful gradients illustrated through an increasing Hellinger distance (red zone), after which we observe a stable training visualized through a decreasing Hellinger distance (green zone), proceeding this we observe convergence whereby the Discriminator provides no meaningful gradients (black zone). As we see in Figure \ref{fig:mnist-3-8-9-track}, convergence was reached at the 38th epoch into training with oscillations around the converged value being attributed to the inherent noisiness of quantum computers and our sampling technique. 

\subsubsection{QuGAN: IBM-Q Evaluation}

\begin{figure}[!htb]
    \centering
    \includegraphics[width=1\linewidth]{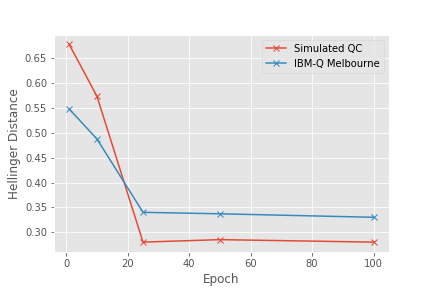}
    \caption{IBM-Quantum v.s. Simulator}
    \label{fig:real_quantum_comp}
\end{figure}

To further evaluate our architecture, 
we run our QuGAN on IBM-Q "Melbourne", a 15-Qubit, 8-Quantum-Volume Quantum Processor through IBM Quantum Experience. Making use of 20 samples per GAN sample, we measure the Hellinger Distance on both the Quantum Computer and the simulator with the same configurations. The measured distances are depicted Figure \ref{fig:real_quantum_comp}, where we illustrate the learning process of our architecture on IBM-Q Melbourne and the simulator. Two key aspects are taken away from this figure, being the noise of quantum computers as illustrated by the inability to converge to lower Hellinger Distances, and the initial better performance using the same parameters due to noise hence covering larger areas of the latent space. 

As can be seen in Figure \ref{fig:real_quantum_comp}, initially the simulator attains worse results than the real quantum computer with values $0.678$ and $0.548$ respectively - a $19.20\%$ difference . However, the simulated Quantum Computer converges closer to the true distribution at a Hellinger Distance of 0.280. This can be attributed to noise, where the simulator is confidently incorrect initially, but becomes more confidently correct as time goes on and can make accurate samples each circuit sampling. In contrast, the IBM-Q results start out with a lower Hellinger Distance, which can be attributed to it being noisy, therefore covering more of the spannable space. IBM-Q converges at a higher Hellinger Distance of 0.337, with the 20.357\% higher distance attributed to the noise of real Quantum Computers. 

%% file: sections/discussion.tex
\section{Conclusions  and Discussion}
In this paper, we proposed QuGAN, a Quantum  Generative Adversarial Network. QuGAN utilizes quantum states to encode classical data and by using Quantum State Fidelity, we make use of a quantum-based loss functions for the Generator and Discriminator. 

The proposed model is evaluated with MNIST dataset. We implement QuGAN with IBM Qiskit and conduct extensive simulations as well as experiments to evaluate QuGAN. It is compared with Tensorflow based classical GANs with different settings, a Qiskit based quantum GAN (Qi-GAN) and Tensorflow-Quantum based GANs in the recent literature (TFQ-GAN). 
The results demonstrate that QuGAN is able to achieve similar performance at meanwhile, reduces 94.98\% of the parameter count compared with classical GANs. Furthermore, it outperforms comparable quantum GANs, Qi-GAN and TFQ-GAN. 

Due to the current limits on quantum computers, it is infeasible to evaluate existing quantum based GANs with ImageNet~\cite{deng2009imagenet} or other large datasets. However, with the rapid evolution of quantum computing, the dimensionality and applicability of QuGANs should be explored as the future work. In addition, further investigation of the effect of deeper quantum neural networks should be considered. QuGAN code is released at \url{https://github.com/yingmao/Quantum-Generative-Adversarial-Network}

